\documentclass[a4paper,fleqn,usenatbib]{mnras}

\usepackage[T1]{fontenc}
\usepackage{ae,aecompl}

\usepackage{graphicx}	
\usepackage{amsmath}	
\usepackage{amssymb}	

\def\gtsima{$\; \buildrel > \over \sim \;$}
\def\simgt{\lower.5ex\hbox{\gtsima}}

\title[The Milky Way globular cluster NGC~7089]{On the physical size of the Milky Way globular cluster NGC~7089 (M~2)}

\author[Andr\'es E. Piatti]{
Andr\'es E. Piatti$^{1,2}$\thanks{E-mail: andres.piatti@unc.edu.ar} \\
$^{1}$Instituto Interdisciplinario de Ciencias B\'asicas (ICB), CONICET-UNCUYO, Padre J. Contreras 1300, M5502JMA, Mendoza, Argentina\\
$^{2}$Consejo Nacional de Investigaciones Cient\'{\i}ficas y T\'ecnicas, Godoy Cruz 2290, C1425FQB,  Buenos Aires, Argentina\\
}

\date{Accepted XXX. Received YYY; in original form ZZZ}

\pubyear{2022}

\begin{document}
\label{firstpage}
\pagerange{\pageref{firstpage}--\pageref{lastpage}}
\maketitle

\begin{abstract}
We study the outer regions of the Milky Way globular cluster NGC~7089 based on
new Dark Energy Camera (DECam) observations. The resulting background cleaned stellar density 
profile reveals the existence of an extended envelope.
We confirm previous results that cluster stars are found out to $\sim$ 1$\degr$ from the
cluster's centre, which is nearly three times the value of the most robust tidal radii estimations. 
We also used results from direct $N$-body simulations
in order to compare with the observations.  We found a fairly good agreement
between the observed and numerically generated stellar density profiles. 
Because of the existence of gaps and substructures along globular cluster tidal tails, we
closely examined the structure of the outer cluster region beyond the Jacobi radius. We
extended the analysis to a sample of 35 globular clusters, 20 of them with observed
tidal tails. We found that if the stellar density profile follows a power law $\propto$ $r^{-\alpha}$,
the $\alpha$ slope correlates with the globular cluster present mass, in the sense that, the
more massive the globular cluster the smaller the $\alpha$ value. This trend is not
found in globular clusters without observed tidal tails. The origin of such a phenomenon
could be related, among other reasons, to the proposed so-called potential escapers or 
to the formation of globular clusters within dark matter mini-haloes.
\end{abstract} 




\section{Introduction}

The detection of tidal tails in Milky Way globular clusters is relevant in the context
of the globular cluster formation environment. Some recent results suggest that the 
absence of tidal tails is related to the fact that globular clusters formed within dark 
matter mini-halos, and role reversal \citep{starkmanetal2019,bv2021,wanetal2021}. 
According to \citet{boldrinietal2020} halo globular clusters formed at or near the centre 
of small dark matter halos still retain to the present day an excess of dark matter above 
the galactic background dark matter. \citet{cg2021} searched for signatures
of dark matter mini-halos in a sample of 25 Milky Way globular clusters. When the
globular cluster sample is matched with the compilation of  globular clusters
with studies of their outer structures by \citet[][see their Table 1]{pcb2020}, we found 
that several globular clusters without tidal tails do not show rising 
velocity dispersion profiles, as expected for those embedded in dark matter
mini-halos \citep{bonacaetal2019}. What we mention above explains in some way why 
the study of the outermost regions of globular
clusters, searching for any kind of extra-tidal feature, is nowadays an active field
of research.

In this work we focus on NGC~7089 (M~2), a globular cluster whose stellar
structure has previously been analyzed from 
different data-sets and methodologies, and for which different results have been obtained.
\citet{grillmairetal1995} studied the outer structure of 12 Milky Way globular clusters,
among them, NGC~7089. They used photographic photometry, and concluded that
the spatial distribution of stars with magnitudes and colours consistent with the
cluster main sequence shows the appearance of tidal tails in their two-dimensional 
surface density map. Although these extra-tidal stars limited the accuracy of
the cluster tidal radius, they determined a \citet{king1966}'s model tidal radius 
of 15.9$\arcmin$.  Later, \citet{dalessandroetal2009} used HST data to trace the
cluster surface density profile and to fit a King model, and obtained
a smaller tidal radius of 9.2$\arcmin$.
\citet{jg2010} also investigated 17 Milky Way globular clusters to identify tidal
tails emerging from them. They employed the Sloan Digital Sky Survey (SDSS)
Data Release 7  \citep{abazajianetal2009}. The density contours 
generated from a colour-magnitude weighted algorithm to map potential cluster 
members resulted similar to the background contours. Contrarily to the expected possible inner
tidal tails aligned with the orbital path of the cluster no large scale features
were detected. \citet{jg2010} derived a mean cluster tidal radius of 11.7$\arcmin$, 
25$\%$ smaller than that of \citet{grillmairetal1995}.

NGC~7089 was also targeted for a search of stellar streams in the outer Galactic halo by
\citet{kuzmaetal2016}, who studied its surroundings from 
Dark Energy Camera and MegaCam data sets. They selected it on the basis of 
a variety of unusual characteristics, which appeared suggestive of an extra-Galactic 
origin. Indeed, the cluster stellar population exhibits a broad dispersion in Fe and
neutron capture elements. The authors concluded that NGC~7089 might plausibly constitute
the stripped nucleus of a dwarf galaxy that was accreted and destroyed by the Milky
Way in the past, although they could not identify the origin of an extended diffuse stellar 
envelope, which embeds it. Such stellar envelope reaches at least $\sim$ 60$\arcmin$,
i.e., several times the tidal radii mentioned above, and has a nearly circular shape,
which decreases in stellar density following a power law with slope $\alpha$ = -2.2$\pm$0.2.

More recently, \citet{deboeretal2019} used {\it Gaia} DR2 data \citep{gaiaetal2016,gaiaetal2018b}
to build radial number density profiles of 81 Milky Way globular clusters. 
Unlike previous studies, they performed a cluster
membership selection from proper motions rather than from magnitudes and
colours.  \citet{deboeretal2019} fitted
the density profiles using a set of single-mass models, among them King's \citep{king1966}
and Wilson's \citep{wilson1975} models, generalized lowered isothermal models, and  a 
spherical potential escapers stitched (SPES) model. For NGC~7089, the model
that best resembles the cluster density profile is the SPES one, with an
associated tidal radius of 19.2$\arcmin$. As can be suspected, the larger
tidal radius could be reflecting the existence of an extended envelope
or tidal tails. We note that \citet{sollima2020} using the same {\it Gaia} DR2 database
did not find any tidal tails. However, long tidal tails associated to NGC~7089 
have been recently detected by \citet{ibataetal2021}, who employed {\it Gaia} EDR3 data and
the {\sc streamfinder} algorithm \citep{mi2018}, and by \citet{grillmair2022}.

In the merger history of the Milky Way, NGC~7089 is associated to the {\it Gaia}-Enceladus 
dwarf galaxy with other 12 globular clusters \citep{kruijssenetal2020}. Among them,
7 have been targeted by studies of their outer regions seeking for hints of tidal tails,
namely: NGC~4147 and NGC~6205 \citep{jg2010}, NGC~6341
\citep{sollima2020}, NGC~6779 \citep{pcb2019}, NGC~6864 \citep{piatti2021d}, NGC~7099 \citep{piattietal2020}, and 
NGC~7492 \citep{naverreteetal2017}. Only NGC~4147 and 7492 exhibit tidal tails. 
We might infer from these results that most {\it Gaia}-Enceladus
globular clusters do not have tidal tails, but they do not have rising velocity dispersions
at large radii as they could have formed within dark matter mini-halos \citep{cg2021}, which
would challenge theoretical models of the formation of globular clusters
\citep[][]{p1984,sz1978}. 

Precisely, Carballo-Bello et al. (see Section 2) embarked in an observing campaign 
of {\it Gaia}-Enceladus globular clusters with the aim of increasing the
number of member clusters with detailed studies of their outer regions, in order to
classify them as clusters with tidal tails, or with extra-tidal envelopes, or without
detectable signatures of extra-tidal structures (NGC~6809 \citep{piatti2021c}; 
NGC~6864 \citep{piatti2021d}; NGC~6981 \citep{piattietal2021}; NGC~7099 \citep{piattietal2020}). 
We focus here on NGC~7089, the last cluster in Carballo-Bello et al.'s sample.
A complete census of the outer regions of  {\it Gaia}-Enceladus globular clusters 
will shed light onto our understanding of whether the formation conditions prevail over times, 
or the orbital history in the Milky Way prevails.

In Section 2 we describe the data sets used, while in Section 3 we use this
data to build the stellar number density profile of NGC~7089.
Section 4 deals with the computation of the cluster orbit from $N$-body simulations,
whose results are compared with those from the observational data and discussed in Section 5.
Although NGC~7089 has been studied extensively, this work contributes to our
knowledge of its outer stellar structure from the focused analysis on 
a cluster colour-magnitude diagram in the SDSS $g$ and $i$ filters,
cleaned from field star contamination using an
independent procedure, and from numerical simulations, which are performed
for the first time. We note that previous works derived tidal radii spanning from 9.2$\arcmin$
up to 19.2$\arcmin$, revealing a real challenge in estimating the cluster's extension. 

\section{Photometric data set}

To explore the outskirts of NGC~7089, we used images obtained with
the Dark Energy Camera (DECam), which is attached to the prime focus of the 
4-m Blanco  telescope at Cerro Tololo Inter-American Observatory (CTIO). DECam is an array of 
62 identical chips  with a scale of 0.263\,arcsec\,pixel$^{-1}$ that provides a 3\,deg$^{2}$ 
field of view
\citep{flaugheretal2015}. The images were taken as part of the observing 
program CTIO 2019B-1003 (PI: Carballo-Bello) and are now of public access. 
NGC~7089  and a comparison field located 5$\degr$ eastward were imaged with 4$\times$600 sec $g$ and 4$\times$400 sec 
$r$ exposures, respectively, under photometric conditions.
Additionally,  observations of  5 SDSS fields at  
different airmass are also available, which were used to derive the
atmospheric extinction coefficients and the transformations between the instrumental 
magnitudes and the SDSS $ugriz$ system \citep{fukugitaetal1996}.

\begin{figure}
\includegraphics[width=\columnwidth]{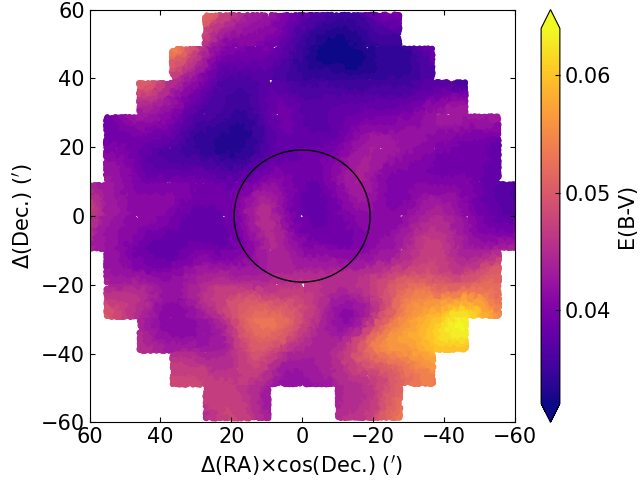}
\caption{Reddening variation across the field of NGC~7089. 
The black circle corresponds to the cluster SPES tidal radius  (19.2$\arcmin$) derived by
\citet{deboeretal2019}.}
\label{fig1}
\end{figure}

\begin{figure*}
\includegraphics[width=\textwidth]{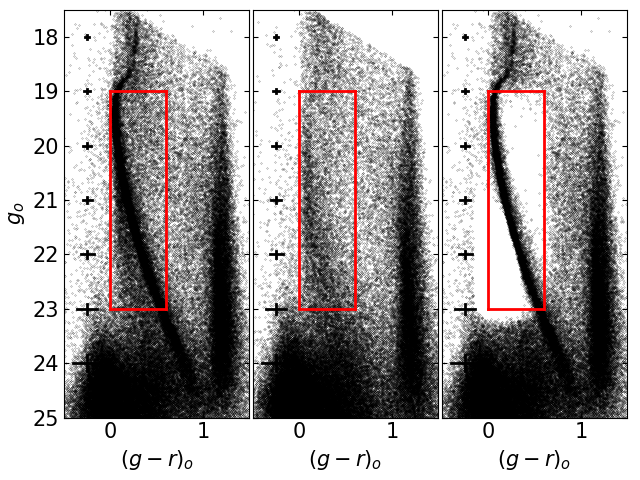}
\caption{Intrinsic CMDs  for all the measured stars in the cluster field
(left panel) and in the comparison field (middle panel). The
cleaned cluster CMD (only for the red delimited region) is shown in the
right panel.
Typical error bars are also indicated.}
\label{fig2}
\end{figure*}

The images were processed following the DECam Community Pipeline \citep{valdesetal2014},
 and the resulting processed images were employed to 
obtain the corresponding photometry using the  \textsc{daophot\,ii/allstar} 
point-spread-function fitting routines \citep{setal90}. Since bad 
pixels, unresolved double stars, cosmic rays, and background galaxies
contaminate our resulting photometric catalogs, we avoided their presence
in our subsequent analysis by imposing the restriction of  keeping positions 
and standardized $g$ and $r$ magnitudes of objects with $|$sharpness$|$
$\leq$ 0.5. Finally, the photometry completeness was estimated from artificial 
star tests, using the {\sc daophot\,ii} routines and the same quality selections as
for observations, as follows: synthetic stars were
added with magnitudes and positions distributed similarly to those of the measured stars
 in each image; their photometry was carried out similarly
as described above; and the resulting magnitudes for the synthetic stars 
were compared with those used to create such stars. While performing the 
artificial star tests, we treated each image without distinguishing between
inner and outer cluster regions. Therefore, on average, the
50$\%$ completeness
level resulted to be 23.4\,mag and 23.3\,mag for the $g$ and $r$ bands, 
respectively. We used
these magnitudes as a reference while choosing the colour-magnitude
diagram regions for building the cluster number density profile.

Fig.~\ref{fig1} shows that there is no
evidence of differential reddening in the NGC~7089 field,
with $E(B-V)$ values retrieved from the 
NASA/IPAC Infrared Science Archive\footnote{https://irsa.ipac.caltech.edu/}
using an input uniform grid of Right Ascension 
and Declination values covering the entire DECam field of view in steps of 1 arcmin, 
respectively. The resulting range of reddening values 
($\Delta$$E(B-V)$ = 0.035 mag) is clearly small. We corrected
the observed $g$ and $r$ magnitudes of each star by using the $E(B-V)$ colour
excesses according to the positions of the stars in the sky. For the
same of the reader, we computed a mean $E(B-V)$ value for the comparison field
of 0.10$\pm$0.02 mag.

We started our analysis by inspecting the cluster colour-magnitude diagram
(CMD) shown in the left panel of Fig.~\ref{fig2}, which includes all
measured stars. As can be
seen, Fig.~\ref{fig2} shows the long cluster main sequence and the 
peculiar contribution of the Milky Way composite star field population, which
can be recognized from the comparison with the star field CMD
(middle panel of Fig.~\ref{fig2}) built from all measured stars in
a DECam field located 5$\degr$ to the East of the cluster's centre.

\section{Stellar number density profile}

A standard approach to build stellar radial profile was used by
\citet{carballobelloetal2014} \citep[see, e.g.,][]{olszewskietal2009,sahaetal2010,roderick15,kuzmaetal2016,kuzmaetal2018}.
However, they did not get rid of
field stars that populate the cluster main sequence too. \citet{deboeretal2019}
employed stars down to {\it Gaia} $G$ = 20 mag with proper motions similar 
to the mean cluster proper motion. We note that main sequence stars are far more numerous than red giants,
so that they are better tracers of low cluster star densities.
We here decided to use main sequences stars, because stars with
smaller masses can be found more easily far away from the cluster main body
\citep{carballobelloetal2012}. For this reason, most of the studies devoted to the 
search for extra-tidal structures have used relatively faint main sequence stars 
\citep[see, e.g.,][]{olszewskietal2009,sahaetal2010}. We also cleaned the
cluster main sequence from field star contamination, so that the distribution in the
sky of the stars that remained unsubtracted represents the intrinsic spatial distribution 
of cluster members.  We describe the process to perform the CMD cleaning in the
subsequent text.

Figure~\ref{fig2} shows a region on the cluster main sequence enclosed
within red boundaries. We selected that portion of the cluster main sequence
to build the respective stellar number density profile, once it is cleaned from field star
contamination.  The route of this cleaning procedure starts by superimposing the comparison 
field CMD on to the cluster CMD; then subtracting from the latter as many stars are in the comparison 
field CMD, choosing those with magnitudes and colours as similar as those of 
the comparison field stars.  The method employed
was introduced by  \citet{pb12}, which was satisfactorily applied elsewhere 
\citep[e.g.,][and references therein]{petal2018,piatti2021d}. The field star decontamination technique
proved to be  successful for clusters projected on to crowded fields and affected by 
differential reddening \citep[see, e.g.,][and references therein]{pft2020}.
In order to clean the selected cluster main sequence region, the method
uses the CMD of the comparison star field as reference and subtracts from the cluster 
CMD the closest star in the magnitude versus colour plane to each field star. 
It assumes that both CMDs share similar
field star properties, namely, magnitude and colour distributions, and stellar density.

The strategy to choose stars to subtract from the  selected cluster CMD region with
magnitudes and colours similars those of the stars in the comparison field CMD
consists in defining boxes centred on the magnitude and colour of each star of the
comparison star field; then to superimpose them on the cluster CMD, 
and finally to choose one star per box to subtract. Ideally, stars with the same
magnitudes and colours of the comparison field stars are desirable. Because
this requirement is difficult to accomplish, stars with magnitudes and colours as
close to those of the comparison field stars are chosen. On the other hand,
because of stochastic effects of field stars, it could not be
straightforward to find a star to subtract in the cluster CMD close to certain 
(magnitude,colour) pair, corresponding to the magnitude and colour of a field star.
In order to assure the subtraction of one star in the cluster CMD per field star
in the comparison field CMD, we started by searching reasonable large areas
in the cluster CMD around the (magnitude,colour) values of each field star.
We started with
boxes with size of ($\Delta$$g_0$,$\Delta$$(g-r)_0$) = (0.25 mag, 0.10 mag)
centred on the ($g_0$, $(g-r)_0$) values of each comparison field star,
in order to guarantee to find a star in the  cluster CMD with
the magnitude and colour within the box boundary. In the case that more than one star 
is located inside that box, the closest one to the centre of 
that (magnitude, colour) box is subtracted. Because magnitudes and colours
of stars in the cluster CMD have uncertainties, the number of stars that fall inside
the boundary of a (magnitude,colour) box depends on whether we consider them.
For this reason,  the magnitude and colour errors of the stars 
in the cluster CMD were taken into account while searching for a 
star to subtract. With that purpose, we allowed the stars in the cluster CMD  
to vary their  positions within an interval of 
$\pm$1$\sigma$, where $\sigma$ represents the errors in their magnitude and colour, 
respectively. We allowed up to a thousand random combinations of their magnitude and colour
errors. Note that the size of the (magnitude,colour) box where the search of a star 
similar in magnitude and colour to a field star is
carried out is not delimited by the
magnitude and colour uncertainties of the stars in the cluster CMD. They are all of the
initial size mentioned above, and inside them the search for
the closest star to its centre is performed.

The aim of the cluster CMD cleaning procedure is to eliminate from it fields stars, so that
only the intrinsic cluster CMD features are visible; the position in the sky of the stars
is not relevant, provided that all of them are within the cluster main body. However, when
extra-tidal structures or tidal tails are searched and large areas reaching far from the
cluster centre are analyzed, the positions in the sky of the subtracted stars play a role.
The results are different if the subtracted stars are all distributed inside that cluster main
body or throughout a larger area. Precisely, the observation of a comparison field
located far from the cluster is aimed at cleaning a reasonable large area around the cluster
main body.
Because of the relatively large extension of the cleaned cluster  area
($\sim$2$\degr$$\times$2$\degr$; see Fig.~\ref{fig1}), we imposed the condition
that the spatial positions of the stars to subtract from the cluster CMD
were chosen randomly. Thus, we avoided spurious overdensities in the
resulting cleaned cluster  area  driven by the subtraction of stars from the cluster
CMD that are located nearly in the same sky regions. In practice,
for each comparison field star, we first randomly  selected the position of a box of 
0.05$\degr$ a side in the cluster  field (see Fig.~\ref{fig1}) where 
to subtract a star. We then looked for a star with ($g_0$, $(g-r)_0$) values within the 
(magnitude, colour) box defined as described above, taking into account the
photometric errors. If no star is found in the selected spatial box, we repeated the 
selection a thousand times, otherwise we enlarged the box size in steps of 
0.01$\degr$ a side, to iterate the process. This alternative is useful
when the number of remaining not subtracted stars with a desire magnitude and colour is
very small, and a star with such magnitude and colour should be find to subtract.
In practice, stars that meet the required magnitude and colour selection are mostly 
find  inside the initial 0.05$\degr$$\times$5$\degr$ boxes randomly distributed
across the DECam field. Note that the CMD cleaning technique for large areas
deals with two main searches, namely; that of stars in the cluster CMD with magnitudes and 
colours similar to those of field stars, and the positions in the sky of those subtracted stars. 
Both iterative searches can be carried out indistinctly of the order of the seach.

The outcome of the cleaning procedure
is a cluster CMD  decontaminated from field stars, i.e., it likely contains only cluster members; their 
spatial distribution relies on a random selection basis. The number of subtracted stars is 
equal to that in the comparison star field, because the subtraction is performed in a star-by-star
basis. However, it could happen that inside a CMD box there is no star to subtract and in this case
the total number of subtracted stars is smaller than that in the comparison field CMD. This can
be checked by counting the number of stars in both the comparison field CMD and those subtracted
from the cluster CMD. In our case, we subtracted the same number of stars. Fig.~\ref{fig2} (right
panel) shows the resulting cluster CMD with the delimited rectangular region decontaminated
from field stars.  By comparing the observed cluster CMD (left panel) with that of the
comparison star field (middle panel), it is readably visible that most of the field star contamination
has been eliminated in the cleaned cluster CMD. 
The cleaned cluster main sequence exhibits the expected shape, with a 
well-defined lower envelope and a blurred upper one caused by the presence of binary stars.
Some isolated stars are visible in the upper-right corner of the devised rectangle,
which can possible come from intrinsic differences between the comparison star field
population and that projected along the line-of-sight of the cluster. NGC~7089 and
comparison field Galactic coordinates are ({\it l},b) = (53.4$\degr$,-35.8$\degr$) and
(56.9$\degr$,-39.8$\degr$), respectively. They do not
affect the analysis of the outermost regions of NGC~7089. On the contrary,
the cleaned CMD regions beyond the cluster main sequence validate the subtraction procedure.
The cleaned CMD also shows that beyond the cleaned region (the red rectangle),
toward bluer and fainter magnitudes,  some stars were subtracted. This is because
we considered the uncertainties in magnitude and colour, so that for a box placed in the
bottom-left corner of the CMD region, stars outside it were selected. The
subtraction of these stars have no effects in our analysis. The most important CMD
region to clean is that along the cluster main sequence, because we use cluster
members to build the cluster density profile.
We executed 1000  times the decontamination procedure, and defined a membership probability $P$ 
($\%$) as the ratio $N$/10, where $N$ is the number of times a star was found 
among the 1000 different outputs. In the subsequence analysis we only kept stars 
with $P$ $>$ 70$\%$.

\begin{figure*}
\includegraphics[width=\textwidth]{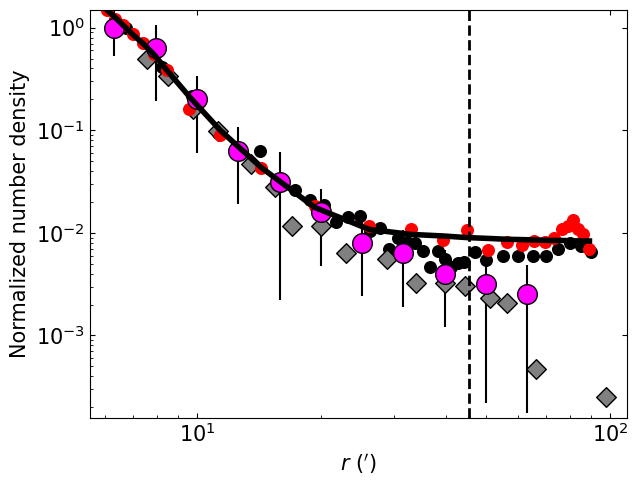}
\caption{Resulting number density profile for stars in the field star decontaminated cluster CMD region.
represented with filled magenta circles. Error bars are also drawn.
The black solid curve  is the  SPES model fitted by
 \citet{deboeretal2019}. 
Grey diamonds represent the stellar radial profile obtained by \citet{kuzmaetal2016};
red filled circles that by \citet{deboeretal2019}, while black filled circles  are obtained from 
the $N$-body experiment. The vertical dashed line is the cluster Jacobi's radius 
 \citep[45.5$\arcmin$,][]{deboeretal2019}.}
\label{fig3}
\end{figure*}

We used stars in the cleaned cluster CMD rectangular region
to build the respective stellar radial profile. In order to do that, we counted the number of stars in annuli
of 1.2$\arcmin$, 2.4$\arcmin$, 3.6$\arcmin$, 4.8$\arcmin$ and 6.0$\arcmin$ wide, and computed their 
average and dispersion.
The resulting radial profile with its respective uncertainties is depicted in Fig.~\ref{fig3}, 
where we included the SPES model fitted by \citet{deboeretal2019}.
 For comparison purposes, we normalized it to the same density at $r$=6.6$\arcmin$.
As can be seen, the present stellar radial profile extend beyond the SPES tidal radius,
which is the largest value that we found in the literature (see Section~1).
The radial profile built by \citet[][their figure 8]{kuzmaetal2016} (grey diamonds in
Fig.~\ref{fig3}) confirms that there is an excess of stars distributed beyond the
SPES tidal radius. Note that the shapes of both radial profiles are remarkably similar.

\subsection{$N$-body simulations}

We also ran a number of direct $N$-body simulations of NGC~7089 in order to compare the observed density
profile with the theoretically expected one for a cluster that has the same mass and is on the same orbit
as NGC~7089. In order to perform the $N$-body simulations, we used the NBODY+P3T code from \citet{arnoldetal2021}. 
This code treats gravitational interactions between stars in close proximity to each other with a fourth order 
Hermite integration scheme and calculates distant gravitational forces using Bonsai \citep{bpz2020}, a 
Barnes-Hut tree code with a standard leap frog scheme. In order to achieve a better performance, we modified the integration routine to
 use a single leap frog scheme, with all gravitational forces calculated using the Tree code. We
chose a critical opening angle of $\theta = 0.4$, a leapfrog integration step size of $\Delta t = \frac{1}{32}$
and a softening length of $\epsilon=0.03$ in $N$-body units (corresponding to about 0.1 pc in physical units or about 1/10th of the core size of NGC~7089)
as the best trade-off between accuracy and speed. This way we were able to perform single simulation of NGC~7089, which would normally take several 
months with a direct $N$-body code like NBODY6, in less than two days.

The general set-up of our simulations follows the ones done in \citet{wanetal2021}. We integrated a point-mass particle backwards
in time along the orbit of NGC~7089 for 2~Gyr with a 4th order Runge-Kutta integrator, using the Milky Way potential by \citet{irrgangetal2013}.
We then set up a star-by-star model of NGC~7089 using the best-fitting model from the grid of $N$-body simulations by \citet{baumgardt2017} and
\citet{bh2018}. This model was then integrated forward in time for 2~Gyr to the present-day position of
NGC~7089 using NBODY+P3T. We slightly varied the mass and size of the initial $N$-body model until we achieved the best-match of our
model cluster to the observed NGC~7089.

We then used the result of the best-fitting $N$-body simulation to construct a stellar
density proifile, similarly to that of Fig.~\ref{fig3}. We then selected from the $N$-body simulation 
results stars
within those mass regimes distributed throughout a region of  2$\degr$$\times$2$\degr$
centred on the cluster as in Fig.~\ref{fig2}, and built the respective stellar density profile
by applying the same recipe as above. The resulting stellar number density profile
is depicted in Fig.~\ref{fig3} with black filled circles . As can be seen,  there is a clear excess 
beyond the SPES radius,  in very
good agreement with the present one and that obtained by \citet{kuzmaetal2016}.

\section{Analysis and discussion}

As far as we are aware, this is the first time that the result of the detailed outer structure
of NGC~7089 built by  \citet{kuzmaetal2016} is confirmed. We used both, independent
data sets and $N$-body simulations to show that NGC~7089 has an extended envelope
beyond the largest tidal radius derived up-to-date. Furthermore, we think that the wide
range of previous tidal radius estimates (from 9.2$\arcmin$ up to 19.2$\arcmin$) is the
result of a combination of a low stellar density structure and the use of data reaching different 
magnitude limits.
\citet{sollima2020} used {\it Gaia} DR2 data sets to explore 5$\degr$ in radius
around NGC~7089. He restricted the search for extra-tidal features to stars brighter than 
{$G$ = 21 mag and applied a 5D mixture modeling technique
and did not find any hints for tidal tails. With 
DECam and MegaCam images \citet{kuzmaetal2016} went even deeper, gaining $\sim$ 
2.5 mag in the limiting magnitude. 
This means that they could compare the radial profile of mainly read giant stars
derived from {\it Gaia} data with that built from much less massive stars.
They did not find tidal tails but a low density
extended envelope following a power law with slope $\alpha$ =-2.2$\pm$0.2.

The existence of tidal tails requires the exploration of larger areas, since they
usually show variations in the stellar density, with the appearance  of gaps, 
spurs, and substructures \citep{odenetal2001,gd2006b,carlbergetal2012,erkaletal2017,deboeretal2020}.
Indeed, the long tidal tails of NGC~7089 identified by \citet{ibataetal2021} and \cite{grillmair2022}
show variations of their stellar densities.
Its leading and trailing
tidal tails turned out to be asymmetric. 
Both tails are nearly aligned along the direction toward the Milky Way centre. We note that the cluster's orbit has an eccentricity of 0.94
and an inclination angle of 84.18$\degr$, so that it describes a trajectory which pass
very close to the Milky Way centre (perigalacticon ($R_{peri}$) = 0.84$\pm$0.06 kpc, apogalacticon 
($R_{apo}$) = 18.80$\pm$0.29 kpc) \citep{baumgardtetal2019}, which explains the tidal tails orientation
\citep{montuorietal2007}. At the present time,
NGC~7089 is nearly midway between its perigalacticon and apogalacticon 
(($R_{apo}$ - $R_{GC}$)/$R_{apo}$ = 0.45) at a galactocentric distance of $R_{GC}$ = 9.68 kpc.
\citet{piattietal2019b}
estimated the ratio of the cluster mass lost by tidal disruption to the total cluster mass
for Milky Way globular clusters ($M_{disp}/M_{ini}$), and derived a value of 0.29 for NGC~7089. 
This value is similar to that of Pal~5 (0.24), which exhibits a $\sim$ 30$\degr$ long
tidal tail  \citep{starkmanetal2019} and its orbit is also inclined (65.13$\degr$) \citep{baumgardtetal2019}. 
However, the latter does not reach the Milky Way bulge
($R_{peri}$ = 17.40 kpc) as NGC~7089 does, and its orbit is not elongated as the NGC~7089's orbit
(Pal~5 orbit's eccentricity=0.17). These similarity and difference could help in disentangling
the origin of tidal tails in globular clusters, and their shapes and substructures, to the light of
two main scenarios, namely, the structure of the Milky Way (e.g., giant molecular clouds, the bar, spiral
arms), or the presence of a population of dark matter subhalos.
At this point, it is worth mentioning that \citet{zm2021} showed for a
sample of nine Milky Way halo globular clusters that the presence of tidal tails is 
subjected to the particular cluster dynamical history.

Bearing in mind that tidal tails does not necessarily emerge from the globular cluster main body
as a steady, smooth stellar density distribution, we focused on the analysis of the extra-tidal features
composed by stars that have escaped the cluster, i.e., those that are placed at distances to the globular cluster's
centre larger than  its Jacobi radius.  We seek any hint for a difference between
globular clusters that exhibit tidal tails with respect to those without them.  For that purpose, we used the 
recent compilation by
\citet[][see their Table~3]{zm2021} of Milky Way globular clusters with studies of their outermost regions. 
They split the globular clusters into three groups, namely: globular clusters with tidal tails (T); those with
extended envelopes (E), and clusters without any signature of extra-tidal features (N). We explored
homogeneously the outer regions of these globular clusters by using the stellar number density profiles 
built by \citet{deboeretal2019} using {\it Gaia} DR2 data, and their adopted globular clusters' Jacobi radii.
We found 36 globular clusters in common between \citet{deboeretal2019} and \citet{zm2021}, and
excluded NGC~5904 from the fit because there is no data outside its Jacobi radius.
Fig.~\ref{fig3} illustrates the Jacobi radius of NGC~7089 with a vertical dashed line and its {\it Gaia}
DR2 stellar density profile depicted with filled red circles, respectively.

We fitted their stellar number density profiles with a power law $\propto$ $r^{-\alpha}$ using points placed 
farther than the globular clusters' Jacobi radii. The resulting $\alpha$ slopes are listed in Table~\ref{tab1},
as well as the number of points ($N$) used during the fit, and the groups to which the globular clusters
belong (T=with tidal tails; E=with extended envelop; N=without extra-tidal signature). As can be seen,
the resulting $\alpha$ slopes are smaller than 1.0, with the sole exception of that for Pal~1 (1.52). The
value obtained for NGC~7089 (-0.65) differs from that of \citet[][2.2]{kuzmaetal2016}, because the latter
considered a much inner region -inside the Jacobi radius-, where a decreasing
stellar number density is apparent (see Fig.~\ref{fig3}). We also note that the $N$-body stellar density 
profile agrees well
with that built by \citet{deboeretal2019} for the whole distance range used. This is not the case of
the stellar number density profiles derived here and by \citet{kuzmaetal2016}, possibly
because of the lack of a better statistics at the boundaries of the observed fields. 

We tried to
correlate these $\alpha$ slopes with different astrophysical properties (e.g., Galactocentric distance,
orbital parameters, structural properties, relaxation times), and 
found that it shows a trend with the present cluster mass \citep{baumgardtetal2019} for globular
cluster with tidal tails. Fig.~\ref{fig4} shows the suggestive correlation, where the symbol size is
proportional to the number of points $N$ used to derive the $\alpha$ values, and the red
line represents a linear least-square fit. As can be seen, the
larger the globular cluster mass, the smaller the $\alpha$ slope; a trend that is not observed in
globular clusters with only extended envelopes or without any extra-tidal feature. These findings
suggests that most massive globular clusters with observed tidal tails can exhibit increasing stellar 
overdensites beyond their Jacobi radii, while their less massive counterparts show decreasing stellar
overdensities as a function of the distance to the globular cluster centre. We also note that no points should have been registered beyond the Jacobi radii 
of globular clusters without extra-tidal features. However, the registered extensions of
globular clusters recovered by \citet{deboeretal2019} from {\it Gaia} DR2 data could differ from the
Jacobi radii used, which were determined by \citet{bg2018}, or else, the group to which
a globular cluster was included could be different. Nevertheless, if the entire globular cluster
sample were treated as having tidal tails, the trend observed for those grouped in the T group
would remain with a larger scatter.

\citet{pcb2020}
and \citet{zm2021} searched for overall kinematic or structural conditions that have allowed some 
Milky Way globular clusters to develop tidal tails, and found that globular clusters behave
similarly irrespective of the presence of tidal tails or any other kind of extra-tidal feature, or the 
absence thereof. As far as we are aware, the present  outcome is the first observed
distinction between globular clusters in the T and E and N groups. Some hints to explain
the correlation between the $\alpha$ slope for distances larger than the Jacobi radius and the 
globular cluster mass (top panel of Fig.~\ref{fig4}) could be found in the proposed so-called 
potential escapers,  which would give rise of the stellar of the density and velocity dispersion 
near the Jacobi radius  \citep{baumgartd2001,kupperetal2010,claydonetal2019}. Such a
rising velocity dispersion at large radii was also suggested by \citet{cg2021} as an observing feature
for globular clusters that could have formed within dark matter mini-halos.

\begin{table*}
\caption{Power law slopes ($\alpha$) of {\it Gaia} DR2 stellar density profiles of Milky Way globulars for distances
beyond their Jacobi radii. The number of points used in the fits ($N$) and the globular cluster
class according to \citet[][T= with tidal tails; E= with extended envelop; N=without extra-tidal signature]{zm2021} 
are also listed.}
\label{tab1}
\begin{tabular}{@{}lccclccclccc}\hline
Cluster & $\alpha$ & $N$ & Group & Cluster & $\alpha$ & $N$ & Group & Cluster & $\alpha$ & $N$ & Group \\\hline
NGC~288		&	0.10			&	2	&	T&NGC~5139	&	-0.26$\pm$0.04	&	35	&	T&NGC~6752	&	-0.16$\pm$0.15	&	8	&	N\\
NGC~362		&	-0.28$\pm$0.15	&	24	&	T&NGC~5272	&	-0.49$\pm$0.93	&	3	&	T&NGC~6809	&	0.04$\pm$0.14	&	13	&	N\\
NGC~1261	&	-0.08$\pm$0.48	&	9	&	T&NGC~5466	&	0.34			&	2	&	T&NGC~6864	&	0.19$\pm$0.38	&	16	&	N\\
NGC~1851	&	0.22$\pm$0.30	&	11	&	T&NGC~5694	&	0.01$\pm$0.14	&	134	&	E&NGC~6981	&	-0.34$\pm$0.31	&	14	&	E\\
NGC~1904	&	-0.02$\pm$0.43	&	12	&	E&NGC~5824	&	0.24$\pm$0.14	&	100	&	T&NGC~7006	&	-0.24$\pm$0.15	&	161	&	N\\
NGC~2298	&	0.54$\pm$0.48	&	5	&	T&NGC~5897	&	0.63$\pm$0.30	&	11	&	N&NGC~7078	&	-0.12$\pm$0.10	&	34	& 	N\\
NGC~2419	&	0.01$\pm$0.30	&	28	&	N&NGC~6101	&	0.43$\pm$0.69	&	3	&	T&NGC~7089	&	-0.65$\pm$0.36	&	12	&	T\\
NGC~2808	&	0.39$\pm$0.17	&	23	&	T&NGC~6205	&	-0.26$\pm$0.22	&	7	&	N&NGC~7099	&	0.68$\pm$0.29	&	5	&	T\\
NGC~3201	&	0.23			&	2 	& 	T&NGC~6229	&	0.07$\pm$0.47	&	8	&	N&NGC~7492	&	-0.83$\pm$0.39	&	3	&	N\\
NGC~4590	&	0.08			&	2	&	T&NGC~6341	&	0.14$\pm$0.34	&	6	&	T&Pal~1		&	1.52$\pm$0.24	&	6	&	T\\
NGC~5024	&	-0.28$\pm$0.78	&	3	& 	T&NGC~6362	&	0.08$\pm$0.30	&	5	&	T&Pal~12		&	-0.10$\pm$0.93	&	5	&	N\\
NGC~5053	&	0.33			&	2	&	E&NGC~6397	&	0.99			&	2	&	T&			&				&		&	\\\hline
\end{tabular}
\end{table*}

\begin{figure}
\includegraphics[width=\columnwidth]{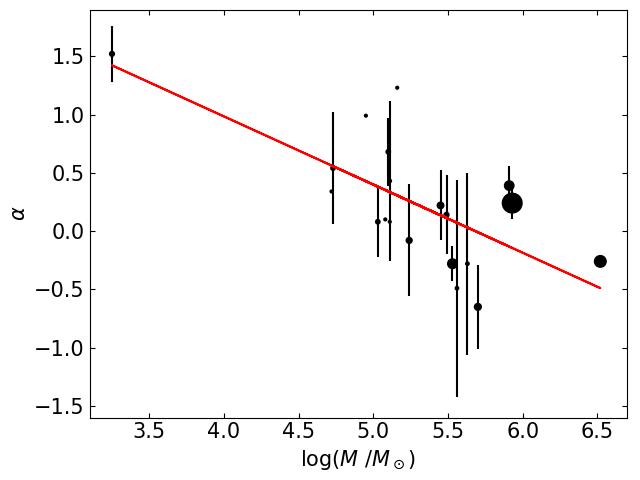}
\includegraphics[width=\columnwidth]{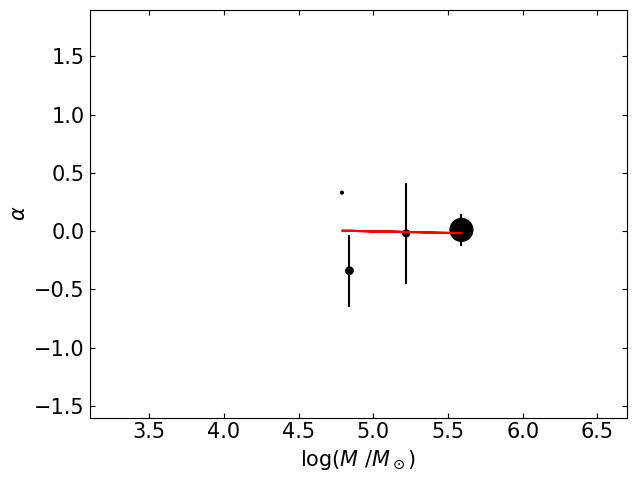}
\includegraphics[width=\columnwidth]{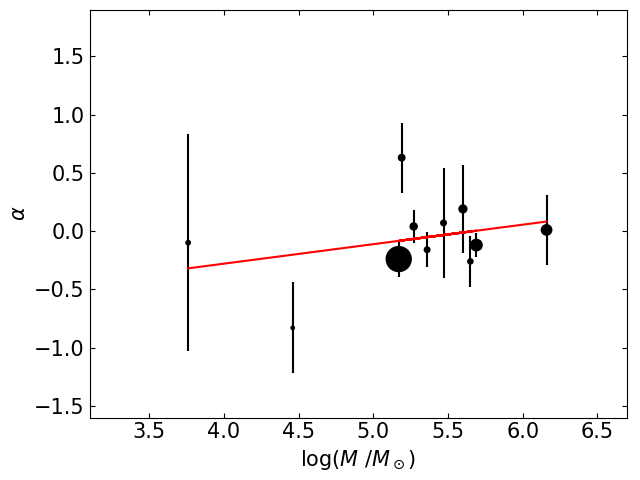}
\caption{Relationship between the power law $\alpha$ slopes
of the outer number density profile (Table~\ref{tab1}) and the present
globular cluster mass. Top, middle, and bottom panels
are for globular clusters in groups T, E, and N, respectively.
The size of the filled circles is proportional to the number
$N$ in Table~\ref{tab1}. The red line represent a linear least-squares
fit.}
\label{fig4}
\end{figure}

\section{Data availability}

Data used in this work are available upon request to the first author.

\section*{Acknowledgements}
I thank the referee for the thorough reading of the manuscript and
timely suggestions to improve it.
I thank the contribution of J.A. Carballo-Bello to an earlier stage of this
project. I warmly thank H. Baumgardt and A. Arnold who carried out the $N$-body experiments and
wrote the respective section of this paper, and made useful suggestions and corrections to the
entire text. I also thank T.J.L. de Boer for making available his catalogue of {\it Gaia} DR2 stellar
number density profiles.

Based on observations at Cerro Tololo Inter-American Observatory, NSF’s NOIRLab (Prop. ID 2019B-1003; 
PI: Carballo-Bello), which is managed by the Association of Universities for Research in Astronomy (AURA)
under a cooperative agreement with the National Science Foundation.

This project used data obtained with the Dark Energy Camera (DECam), which was constructed by the 
Dark Energy Survey (DES) collaboration. Funding for the DES Projects has been provided by the US 
Department of Energy, the US National Science Foundation, the Ministry of Science and Education of Spain, 
the Science and Technology Facilities Council of the United Kingdom, the Higher Education Funding Council 
for England, the National centre for Supercomputing Applications at the University of Illinois at 
Urbana-Champaign, the Kavli Institute for Cosmological Physics at the University of Chicago, centre for 
Cosmology and Astro-Particle Physics at the Ohio State University, the Mitchell Institute for Fundamental 
Physics and Astronomy at Texas A\&M University, Financiadora de Estudos e Projetos, Funda\c{c}\~{a}o 
Carlos Chagas Filho de Amparo \`{a} Pesquisa do Estado do Rio de Janeiro, Conselho Nacional de 
Desenvolvimento Cient\'{\i}fico e Tecnol\'ogico and the Minist\'erio da Ci\^{e}ncia, Tecnologia e Inova\c{c}\~{a}o, the Deutsche Forschungsgemeinschaft and the Collaborating Institutions in the Dark Energy Survey.
The Collaborating Institutions are Argonne National Laboratory, the University of California at Santa Cruz, the University of Cambridge, Centro de Investigaciones En\'ergeticas, Medioambientales y Tecnol\'ogicas–Madrid, the University of Chicago, University College London, the DES-Brazil Consortium, the University of Edinburgh, the Eidgen\"{o}ssische Technische Hochschule (ETH) Z\"{u}rich, Fermi National Accelerator Laboratory, the University of Illinois at Urbana-Champaign, the Institut de Ci\`{e}ncies de l’Espai (IEEC/CSIC), the Institut de F\'{\i}sica d’Altes Energies, Lawrence Berkeley National Laboratory, the Ludwig-Maximilians Universit\"{a}t M\"{u}nchen and the associated Excellence Cluster Universe, the University of Michigan, NSF’s NOIRLab, the University of Nottingham, the Ohio State University, the OzDES Membership Consortium, the University of Pennsylvania, the University of Portsmouth, SLAC National Accelerator Laboratory, Stanford University, the University of Sussex, and Texas A\&M University.










\bsp	
\label{lastpage}
\end{document}